\input psfig.tex
\documentstyle[12pt,aasms4]{article}

\lefthead{Silk \& Spaans}
\righthead{High Redshift CO Lines}

\begin{document}

\title{Molecular Lines as Diagnostics of High Redshift Objects}

\author{Joseph Silk}

\affil{Departments of Astronomy and Physics, University of California, 601 Campbell Hall, Berkeley, CA 94720}

\author{Marco Spaans}

\affil{Department of Physics \& Astronomy, The Johns Hopkins University,
3400 North Charles Street, Baltimore, MD 21218-2686}

\begin{abstract}

Models are presented for CO rotational line emission by high redshift
starburst galaxies. The influence of the cosmic microwave background on the
thermal balance and the level populations of atomic and molecular species is
explicitly included. Predictions are made for the observability of starburst
galaxies through line and continuum emission between $z=5$ and $z=30$. It is
found that the Millimeter Array could detect a starburst galaxy with
$\sim 10^5$ Orion regions, corresponding to a star formation rate of
about $30 \rm M_\odot yr^{-1}$, equally well at $z=5$ or $z=30$ due to the
increasing cosmic microwave background temperature with redshift.
Line emission is a potentially more powerful probe than dust continuum
emission of very high redshift objects.

\end{abstract}

{\it subject headings}: cosmology: theory - galaxies: starburst -
galaxies: evolution - ISM: molecules

\section{Introduction}

Searches for CO emission from cosmological objects
have had some success. Examples include 
the IRAS source F10214+4724 at $z= 2.29$ (Solomon, Downes and Radford 1992),
the clover-leaf galaxy at $z=2.6$ (Barvainis et al.~1994), and a quasar
at $z=4.69$ (Omont et al.~1996).
These searches establish that large amounts of molecular gas are present at
high redshifts. This is to be expected in the light of the recent
detections of high redshift Lyman break galaxies, e.g.~in the Hubble Deep
Field, which are actively forming stars at $z\sim 3-4$ (Steidel et al.~1996).
Since star formation is ultimately
driven by the collapse of cold molecular clouds, the occurrence of active star
formation must be reflected by the physical structure of the interstellar
medium (ISM). The detection of molecular gas and dust at high redshift
therefore provides an excellent probe of the stellar processes occurring in
cosmological objects. In fact, the metallicity and physical state of the high
redshift ISM provides indirect constraints on the star formation rate and 
hence on models of galaxy formation.

In the next decade, instruments will come on line to explore the infrared and
millimeter regions of the spectrum with the goal of detecting objects at very
high redshift: NGST, FIRST, and the MMA. The latter will have the ability to
detect emission lines fluxes at the milliJansky level around wavelengths of a
few millimeters, and will be ideal for a
search for highly redshifted molecular lines, in particular for CO lines.
In this work the excitation of the CO molecule will be investigated in the
presence of a warm Cosmic Microwave Background (CMB) and the subsequently
altered thermal balance. The aim is to determine which molecular lines are
best suited for the detection of star-forming primordial galaxies, and to
assess up to which redshift such measurements are feasible with the planned
next generation of observatories.

\section{Cosmology}

Cosmology does not provide robust predictions for the redshifts of the first
galaxies to form. Large-scale structure measurements directly constrain the
power spectrum of primordial density fluctuations on scales of
$\sim 5 \rm Mpc$, or larger, and information on smaller scale power comes
from relatively indirect constraints, such as the abundance at high redshift
of damped Lyman alpha absorption clouds. Theory can provide sufficient power
to form the first moderately massive galaxies at a redshift as high as 10,
and in rare instances at even higher redshift, although popular theories of
the cold dark matter variety with nearly scale-invariant primordial power
spectra would be hard-pressed to produce many massive galaxies at $z>10$.
However, observations are the ultimate arbiter of when galaxies form, and
massive galaxies or protogalaxies certainly appear to be present at $z\sim 5$.

Moreover it is generally  believed that spheroids are the oldest components
of galaxies, and formed in an ultraluminous starburst. Such single starburst
models for spheroids are well established by late-time spectral modelling,
and at early epochs the lack of success in optical searches for ultraluminous
forming galaxies has motivated the inference that spheroids form in
ultraluminous dust-shrouded starbursts at high redshift.
Observations of nearby ultraluminous infrared galaxies indeed reveal many
features in common with the presumed properties of protospheroids. This
includes the occurrence of a major merger, a de Vaucouleurs profile in the
near infrared light, a gas surface density and scale size that are comparable
to the stellar characteristics of a spheroid, and a star formation rate of
several hundreds of solar masses per year (Zepf and Silk 1996;
Spaans \& Carollo 1997; and references therein).
We note parenthetically that there is however a rival class of theories which
maintains that most of the stars in spheroids formed in many smaller
mergers (Kauffmann \& Charlot 1997).

Our best prototype of such luminous starbursts comes from infrared
observations, which demonstrate that most of the stellar luminosity is
re-emitted by dust in the far infrared (FIR). However, forming spheroids will
be extremely gas-rich, and one would expect much of the radiation to be
emitted as molecular lines.
The cosmic microwave background, we will now show, has a profound influence
on the populations of the rotational levels responsible for the strongest
lines, and leads to a signature that is potentially detectable out to very
large redshift.

\section{The Model}

A generic starburst galaxy will be asssumed to consist of $3\times 10^5$
Orion regions with the model as described in Hogerheijde
et al.~(1995) and Jansen et al.~(1995). The strength of the illuminating
OB association $\theta^1$ Ori is $I_{\rm UV}=5\times 10^4$ in units of the
average interstellar radiation field (Draine 1978). The numerical code
developed by Spaans (1996) is used
to solve the chemical and thermal balance equations including the effects
of dust attenuation and radiative transfer in the cooling
lines of O, C$^+$, C, Si, Si$^+$, Fe, Fe$^+$, and CO. The main features of the
code include 1) a three-dimensional inhomogeneous density distribution, 2)
heating and cooling processes as described in Tielens \& Hollenbach (1985a),
3) a chemical network consisting of 165 species and 648 reactions, and 4) a
Monte Carlo treatment of the above radiative transfer effects which
incorporates the self-shielding transitions of H$_2$ and CO.

The inhomogeneous density distribution includes a low density medium of
$6\times 10^4$ cm$^{-3}$ with high density clumps of $2\times 10^6$ cm$^{-3}$
and is taken from the Hogerheijde et
al.~model. The clumps contain $\approx 10$\% of the material, but fill only
$\approx 0.3$\% of the volume.
This distribution provides a good fit to the extensive observations
available for the Orion Bar region. For the CO molecule, the
level populations are determined in statistical equilibrium with explicit
inclusion of the redshift dependent CMB. All the features of the
Galactic Orion region are retained, but various metallicities are considered
because these strongly influence the thermal balance and chemical equilibrium.
Further details of cosmological
applications of the code are provided in Spaans \& Norman (1997).

The Orion molecular cloud complex has about 
$5\times 10^4$ solar masses in Orion A. Only about ten percent of this mass
is in the vicinity of the star-forming region, the Orion Nebula Cluster,
centered on the Trapezium (Bally et al.~1987). The total FIR
(40-300 $\mu$m) luminosity of the Orion star forming region is about $10^5$
solar luminosities. This results in a ratio of far-infrared  luminosity to
molecular gas mass of about $20 \rm L_\odot/M_\odot$.
The integrated intensity of all the FIR
fine-structure lines is about $5\times 10^{-2} \,\rm erg \, cm^{-2} s^{-1} sr^{-1}$ at
the Trapezium. The integrated FIR continuum intensity at this position is
about $12 \,\rm erg\, cm^{-2} s^{-1} sr^{-1}$. This yields a photo-electric
heating efficiency of about $4\times 10^{-3}$ (Tielens \& Hollenbach 1985b).

The main contribution to the high-J CO lines comes from the so-called Orion
Bar region, rather than the molecular cloud, where densities reach
$10^6$ cm$^{-3}$, sufficient to excite the $J<10$ lines in terms of their
critical density. This region is less than 1 pc in
physical size and will remain beam diluted at any reasonable ($z<100$)
redshift for the MMA with its beam size of 0.1$'$. In estimating the
observed flux, we have incorporated  the well known redshift dependences of
surface brightness and angular size diameter.


\section{Results and Discussion}

The total molecular gas mass in Orion is about $2\times 10^5\,\rm M_\odot$.
The star formation rate in Orion is probably about
$3\times 10^{-4} \rm M_\odot \, yr^{-1}$, according to recent observations of
the Orion Nebula Cluster (Hillenbrand 1997).
Hence our putative protogalaxy containing $3\times 10^5$ Orions is forming
stars at a rate of $90\rm \, M_\odot \, yr^{-1}$ from a gas reservoir of about
$6\times 10^{10} \, \rm M_\odot$.
These numbers are merely meant to be representative for a protospheroid of
modest mass, amounting
to only 10 percent or so of the characteristic stellar mass $M_\ast$ 
(as defined by the galaxy luminosity function) of an elliptical
galaxy if the star formation efficiency of the gas is about 50 percent. 
Our results can of course be trivially rescaled. The local ratio of
FIR luminosity to molecular gas mass of about
$20 \rm L_\odot/M_\odot$ is in the range associated with luminous starbursts
(Sanders, Scoville \& Soifer 1991). This ratio will vary with location.
A more typical value is about half of this, for then
the FIR luminosity is inferred to be about $6\times 10^{11} \rm L_\odot$
and scales appropriately with the global star formation rate for the Milky Way.
Hence our model of $3\times10^5$ Orions
matches in luminosity, molecular mass, and star formation rate what would be
expected from a luminous starburst.

The column-averaged population distribution of CO is shown in Figure 1a
(top panel) as a function of redshift.
Line intensities averaged over the source are shown in Figure 1b.
The peak in the level population shifts with increasing redshift
to J=5-6, with critical densities $n_{\rm c}\sim 10^5$ cm$^{-3}$ and
excitation energies of $\sim 80-100$ K for the corresponding
transitions, at $z\sim 30$, before becoming thermalized by the CMB. Note that
the line luminosities increase as the CMB becomes hotter because
the level excitation increases with increasing J
until a CMB temperature $T\sim 90 \rm K$ is attained and higher J lines cannot
be pumped in an Orion-like environment due to their high critical density.

The predicted line fluxes are shown in Figure 2. Because of the enhancement
of line fluxes by the CMB, we find the remarkable result that beyond $z\sim 5$,
the predicted line intensities are almost independent of redshift.
Solomon, Radford and Downes (1992) previously noted that the CO(J=$3-2$) line
is always comparable in strength to the CO(J=$1-0$) line up
to $z\sim 2$ due to the warmer microwave background.  We find that as
higher rotational levels are populated at higher redshift, 
one can
measure a starburst at $z\sim 30$ as easily as at $z\sim 5$. In fact, the
measurement may be even easier since the emission peaks at longer wavelengths.
The upper and lower grey curves in Figure 2 indicate the range in fluxes for
Orion-like regions with metallicities which are 4 times higher and 4 times
lower than solar. The overall effect is a factor of a few, which indicates
that the effect of the CMB on the CO line intensities is robust.

All lines of interest are in the millimeter range, and the milliJansky fluxes
predicted for our $3\times 10^5$ Orion model are within the capability of the
MMA to eventually be measurable. Of course, one would have to search 
a considerable amount of sky. If the duration of a starburst is
$\sim 10^8\alpha$yr, and $0.01\beta$ is the fraction of early-forming
spheroids of mass above $0.1 M_\ast$ relative to present-day ellipticals,
one would at most expect $\sim 100 \alpha\beta$ per sq.~deg.~at $z\sim 5$,
and an order of magnitude fewer at $z\sim 30$.

It is of interest finally to compare the FIR emission with our
predicted line fluxes.
In Figure 3, we show the continuum fluxes estimated for the Orion dust model.
As one proceeds to redshifts above $\sim 5$ the dust emission becomes
inreasingly harder to detect relative to the line emission.
At 100 $\mu$m, the typical continuum flux is about 1 mJy for the $z=5$ model.
It will be a challenge even for FIRST to detect such a weak signal,
requiring days of integration. Conversely,
the MMA sensitivity limit for spectral lines is expected to be about 1mJy,
as compared to our predicted fluxes of several mJy for our model of what is
only in effect a modest starburst by ultraluminous infrared galaxy standards.

We comment in closing that the main emphasis has been on the rotational lines
of CO because
of the fortunate energy level spacing for increasing redshifts. Other
molecular species such as CS, HCO$^+$ and HCN will be bright emitters as well,
but do not couple as favorably with the CMB. Therefore, their fluxes will be
strongly reduced by cosmological redshift effects and not easily detectable at
high redshift.

MS acknowledges with gratitude support of NASA grant NAGW-3147 from the Long
Term Space Astrophysics Research Program. The research of JS has been
supported in part by grants from
NASA and NSF. JS also acknowledges with gratitude the hospitality
of the Physics Department at the Johns Hopkins University 
as a Bearden Visiting Professor, and the Institute of Astronomy
at Cambridge as a Sackler Visiting Astronomer.

\newpage
\clearpage

\begin{figure}
\caption{Top Panel: Redshift dependence of the column averaged CO level
populations for an Orion region
at various redshifts. Note the shift in the peak of the rotational level
excitation for higher values of the CMB temperature. Bottom Panel:
The corresponding line intensities averaged over the source.}
\end{figure}

\begin{figure}
\caption{Redshift dependence of the CO emission spectrum for a starburst
galaxy containing $3\times 10^5$ Orion regions. The grey lines indicate the
range in line intensities resulting from metallicities equal to 4 and 1/4
times solar. Note the rough constancy of the line luminosity with redshift.}
\end{figure}

\begin{figure}
\caption{Redshift dependence of the dust spectrum for the same model
as in Figure 2.}
\end{figure}

\setcounter{figure}{0}

\newpage
\begin{figure}
\centerline{\psfig{figure=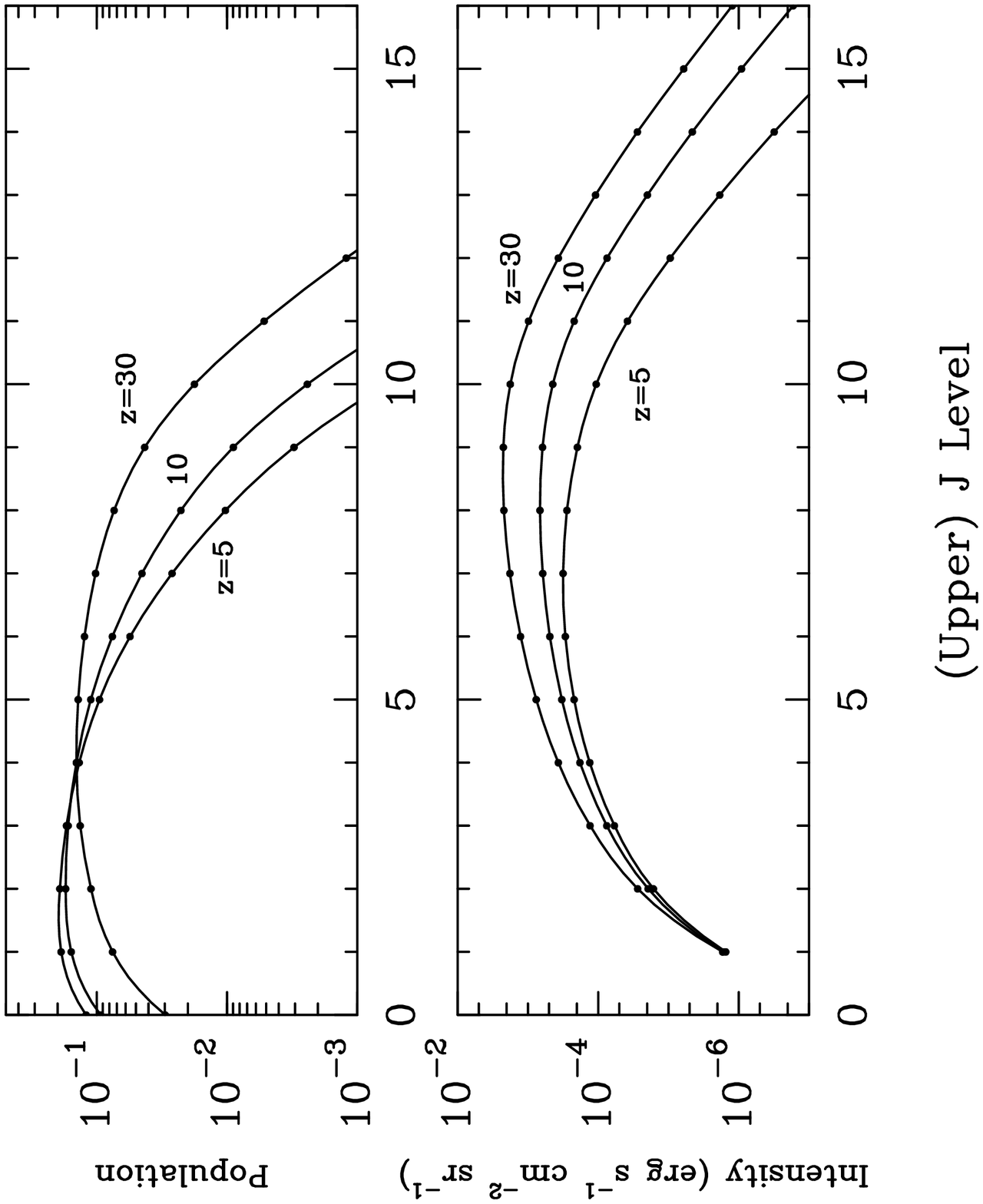,width=6in}}
\caption{\label{figure1}}
\end{figure}

\newpage
\begin{figure}
\centerline{\psfig{figure=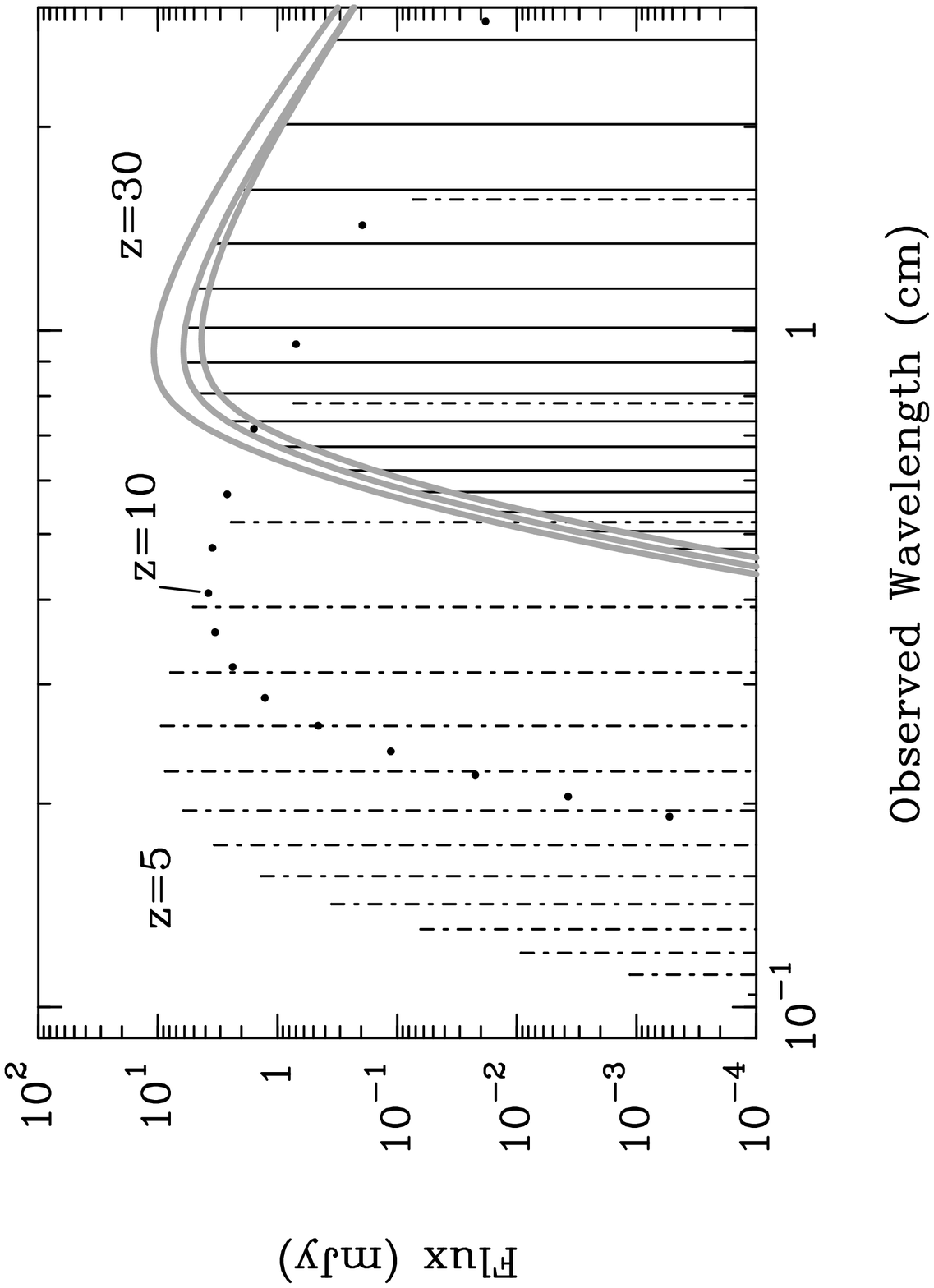,width=6in}}
\caption{\label{figure2}}
\end{figure}

\newpage
\begin{figure}
\centerline{\psfig{figure=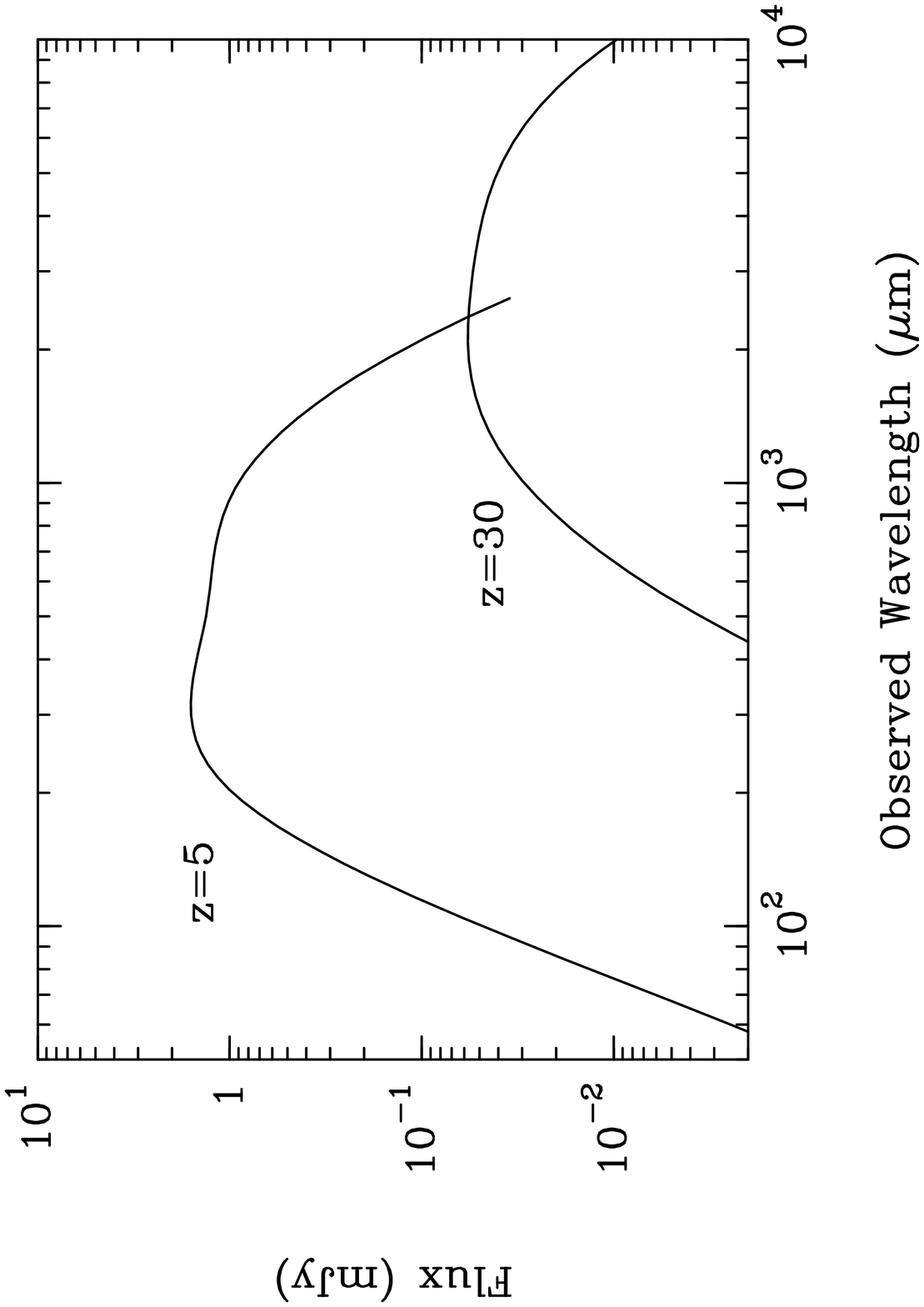,width=6in}}
\caption{\label{figure3}}
\end{figure}

\end{document}